\documentclass[runningheads]{llncs}
\usepackage{graphicx}
\usepackage{color}
\usepackage[table]{xcolor}
\usepackage{colortbl}
\usepackage{comment}
\usepackage{hyperref}

\begin{document}
\title{Federated Learning for Open Banking}

\author{Guodong Long*\inst{1}\orcidID{0000-0003-3740-9515} \and
	Yue Tan\inst{1}\orcidID{0000-0001-8369-2521} \and \\
	Jing Jiang\inst{1}\orcidID{0000-0001-5301-7779}\and
	Chengqi Zhang\inst{1}\orcidID{0000-0001-5715-7154}}
\authorrunning{Guodong Long et al.}

\institute{Australian Artificial Intelligence Institute (AAII), \\ 
            Faculty of Engineering and IT (FEIT), \\ 
            University of Technology Sydney (UTS), Australia\\
            \email{guodong.long@uts.edu.au, yuetan031@gmail.com,\\
            jing.jiang@uts.edu.au, chengqi.zhang@uts.edu.au}}
\maketitle              
	
\begin{abstract}
	Open banking enables individual customers to own their banking data, which provides fundamental support for the boosting of a new ecosystem of data marketplaces and financial services. In the near future, it is foreseeable to have decentralized data ownership in the finance sector using federated learning. This is a just-in-time technology that can learn intelligent models in a decentralized training manner. The most attractive aspect of federated learning is its ability to decompose model training into a centralized server and distributed nodes without collecting private data. This kind of decomposed learning framework has great potential to protect users' privacy and sensitive data. Therefore, federated learning combines naturally with an open banking data marketplaces. This chapter will discuss the possible challenges for applying federated learning in the context of open banking, and the corresponding solutions have been explored as well. 
		
\end{abstract}
\section{Introduction}
    As a subspecies to the open innovation \cite{chesbrough2014new,chesbrough2003open} concept, open banking is an emerging trend in turning banks into financial service platforms, namely banking as a service. From a financial technology perspective, open banking refers to:\cite{open2018open} 1) the use of open application programming interfaces (APIs) that enable third-party developers to build applications and services around the financial institution, 2) greater financial transparency options for account holders ranging from open data to private data, and 3) the use of open-source technology to achieve the above.\cite{brodsky2017data} Open banking can be naturally evolved into a new ecosystem of data marketplaces where participants can buy and sell data. 
    
    As stated by McKinsey \& Company \cite{brodsky2017data}, open banking could bring benefits to banks in various ways, including better customer experience, increased revenue streams, and a sustainable service model for under-served markets. Open banking will form a new ecosystem for financial services by sharing banking data across organizations and providing new services. However, there are inherent risks in sharing banking data, which is sensitive, privacy-concerned, and valuable. It is critical to developing processes and governance underpinning the technical connections. Moreover, the European Union's General Data Protection Regulation (GDPR) \cite{gdpr2016eu} enforces organizations to pay great attention when sharing and using customers' data.

    In the new financial ecosystem, a number of small and medium-sized enterprises will provide novel applications using artificial intelligence (AI) technology. Intelligent applications are already driving a dramatic shift in how financial institutions attract and retain active customers. In recent times AI has become widely applied to review small loan applications fast and automatically. In this AI wave, the machine learning (ML) model with a deep neural network architecture has been a huge success in many financial applications. Imagine the open banking scenario in the near future, with a vast amount of customer data derived from various financial service providers that can be integrated and utilized to train a comprehensive AI model superior to all existing models. Federated learning (FL) is a decentralized ML framework that is able to collaboratively train an AI model while preserving user privacy. It is naturally suited to distributed data ownership settings in an open banking scenario.
    	
    In the context of open banking, federated learning needs to be adapted and enhanced to solve a few practical challenges, such as broad heterogeneity across users, limited times to access personal data, narrow scope of one user, and managing incentives for data contributors. In the following sections, we will briefly introduce applications of open banking, and then discuss the practical challenges with corresponding techniques that present solutions.

\section{Applications of open banking}
\subsection{Open innovation and open banking}
     \paragraph{Open innovation} is ``a distributed innovation process based on purposively managed knowledge flows across organizational boundaries, using pecuniary and non-pecuniary mechanisms in line with the organization's business model" \cite{chesbrough2014new}. The flows of knowledge may involve various ways to leverage internal and external resources and knowledge. Open banking is a kind of open innovation in the banking industry. By leveraging both internal and external knowledge, many innovative applications will emerge to benefit the whole finance industry, including both banks and third-party companies. 
    
    \paragraph{Open banking} \cite{brodsky2017data} can be defined as a collaborative model in which banking data is shared through APIs between two or more unaffiliated parties to deliver enhanced capabilities to the marketplaces. The potential benefits of open banking are substantial including customer experience, revenue, and new service models. It will also improve the finance industry by including more small and medium-sized players with innovative ideas and fine-grained service for different segmentations of customers. In addition to well-known players like Mint, examples include alternative underwriters ranging from Lending Club in the United States to M-Shwari in Africa to Lenddo in the Philippines, and payments disruptors like Stripe and Braintree.
    
    \paragraph{The United Kingdom's} The Competition and Markets Authority issued a ruling in August 2016 that required the UK's nine biggest UK retail banks at that time – HSBC, Barclays, RBS, Santander, Bank of Ireland, Allied Irish Bank, Danske Bank, Lloyds, and Nationwide – to allow licensed startups direct access to their data down to the level of account-based transactions. By August 2020 there were 240 providers regulated by the Financial Conduct Authority enrolled in open banking. They include many providers of financial apps that help manage finances as well as consumer credit firms that use open banking to access account information for affordability checks and verification.
    
    \paragraph{Australia} launched an open banking project on 1 July 2019 as part of a Consumer Data Rights (CDR) project of the Australian Treasury department and the Australian Competition and Consumer Commission (ACCC). The CDR is envisaged to become an economy-wide system that will enable the safe and secure transfer of consumer data. CDR legislation was passed by the Australian Parliament in August 2019. From 1 July 2020 Australia's bank customers have been being able to give permission to accredited third parties to access their savings and credit card data. This enables customers to search for better deals on banking products and to track their banking in one place.
    
    \paragraph{China's} state media reports that China’s financial authorities plan to launch new regulations and policies for the open banking sector and open APIs. Economic Information Observer said that Chinese authorities will “unveil policies in relation to the regulation of open API’s and accelerate the formulation of regulatory standards in relation to open banking in China.” Chinese authorities will also strengthen the regulation of client-end software provided by financial institutions, and expand filing for mobile financial apps from trial areas to the whole of the country. Some Internet giants have already dived into this new trend. For example, Tencent Cloud and WeBank collaborate to launch a Fintech Lab to explore open banking in China. PwC China also has an investigation report with a high-level design for the open banking ecosystem in China \cite{zhang2019@open}.

\subsection{Open banking related applications}
    Open banking has significantly advanced along various pathways \cite{pymnts2020b2}. They include enhancing intelligent applications in existing areas of banking, such as fraud detection, assessment of loan or credit card, customer retention, and personalized service. Below we will introduce the recent developments in open banking related applications worldwide.

    \paragraph{Payment management:}
    In February 2020 payments organization Nacha announced a rollout of an online platform, namely Phixius, that integrates technology, rules, and participants to exchange payment-related information across ecosystem participants. It is associated with the current reliance on proprietary bilateral data-sharing agreements, which limits broader efficiency. The platform is intended to enable companies to exchange payment-related information to improve fraud protection, automate manual processes, and improve customer experiences. 
    Moreover, digital payments company Payrailz announced a partnership with a credit union and FinTech collaborator, namely Constellation Digital Partners, to develop an elevated joint payment solution for credit union clients.

    \paragraph{API Integration:}
    The UK's Starling Bank has been expanding its Business Marketplaces for small businesses by integrating multiple ecosystem participants including Mortgage lender Molo, a freelance career management portal, and accounting system PayStream. These participants can access Starling Bank account data via an API, and it enables the financial institution to connect businesses with integrated solutions. Moreover, in February 2020 the Commonwealth Bank of Australia launched a free app that enables small businesses to consolidate data from various platforms, such as Xero and Google Analytics. This solution, Vonto, is designed to offer transparent and quick insights, connecting small business owners with 10 “key insights” each morning, including cash flow, social media engagement, and website traffic. The UK's Simply Asset Finance collaborates with open banking platform AccountScore, to enable Simply to wield an API to unlock borrower data for deeper underwriting capabilities. The companies said they will use the rich data set to assist lending decisions.

\subsection{Federated learning for open banking}
    \paragraph{Federated learning} is a decentralized machine learning framework that can train a model without direct access to users' private data. The model coordinator and user/participant exchange model parameters that can avoid sending user data. However, the exchanging of model parameters, or gradients in machine learning terminology, may cause data leakage \cite{gdpr2016eu}. Therefore, differential privacy \cite{dwork2008differential} technology is essential for federated learning to protect privacy from gradient-based cyber-attack \cite{abadi2016deep}.

    \paragraph{Data sharing} is the key idea in open banking. As there are inherent risks during data sharing, it is critical to develop processes and governance underpinning the new trend \cite{brodsky2017data}. Customers are more likely to not sell the data, but to use the data to train a model in the local device. Moreover, as required by the GDPR, the shared data for a particular financial service, e.g. credit card application, cannot be used for another purpose, e.g. model training. It is therefore a natural solution to integrate federated learning with an open banking data marketplaces.

    \paragraph{Privacy concerns} are top priority in federated learning. In most cases,  malicious attackers pretend to be a model coordinator of federated learning and can then use gradient-based privacy attack methods to guess what user data look like, to cause privacy leakage. Therefore, differential privacy, secure aggregation \cite{bonawitz2017practical}, and homomorphic encryption are widely used methods for data protection in the federated learning framework. This chapter does not discuss details of privacy-preserving techniques and data encryption as they have been well studied in many literature reviews \cite{mirshghallah2020privacy,shokri2015privacy}. 

	\paragraph{Incentive management} is a practical challenge in open banking. Studies of this problem are two-fold: 1) how to incentivize data owners to participate in federated learning by contributing their data, and 2) how to measure each participant's contribution. Different forms of incentives are possible, such as user-defined utility and money-based rewards, and have been well discussed in the literature \cite{kang2019incentive,khan2019federated,yang2019federatedbook,zhan2020learning}.
	
	\paragraph{Data heterogeneity} is an inherent challenge in a large-scale system across many organizations. In open banking, the user may come from different banks with different feature spaces. The same user may have different pieces of information in multiple banks. To utilize these data in federated learning, the major issue is to align the heterogeneous structured data via horizontal/vertical federated learning and federated transfer learning \cite{yang2019federated}.
	
	\paragraph{Statistical heterogeneity} is caused by the diverse nature of user behaviors. Each user's data may vary in its hidden distribution. This kind of heterogeneity challenge widely exists in a large-scale machine learning system. 
	
	\paragraph{Model heterogeneity} is the scenario that different participants may choose and run a model with personalized model architectures. It is critical to solving the problem of how a central server can aggregate information across participants with heterogeneous models.
	
	\paragraph{Charging by access times} is possible in federated learning. Access to a user's profile is not unlimited and could be charged by times. This raises the question of how the model trainer can complete the training process by accessing a user's profile in a few rounds, namely federated learning with few-shot round communications.
	
	\paragraph{Only positive labels} arise because each user usually only has one-class data while the global model trains a bi-class or multi-class classifier. For example, if we want to train a fraud detection model, we find that most users only have non-fraud data. Training on participants' data with only positive labels is a challenge known as a one-class problem. Aggregating these one-class classifiers is also a new challenge in federated learning.
	\\
	\\
	In the following sections, we discuss the statistical heterogeneity, model heterogeneity, access limits, and one-class problems that are rarely discussed in other places.

\section{Problem formulation}    
	The learning process of federated learning is decomposed into two parts that occur in different places: server (coordinator) and nodes (participants). These two parts are linked to each other via a specifically designed mechanism. In particular, the participant $i$ can train a local model $h_i$ using its own dataset $D_i = \{(x_{i\cdot}, y_{i\cdot})\}$. The model $h_i$ is initialized by a globally shared model parameter $W$ which is then fine-tuned to a new local model with parameters $W_i$ using the data from node $i$.
	
	It is proposed that the coordinator in federated learning can learn a global model controlled by $W$ that could be shared with all participants on distributed nodes. Through a few rounds of communication, the global model has been gradually improved to better suit all participants, and the final global model is an optimal solution that could directly be deployed on each participant for further use. In particular, the optimal global model is expected to minimize the total loss of all participants, and it is defined as below.
	\begin{equation}
	    \sum_{i=1}^n p_i \cdot L(D_i,W) = \sum_{i=1}^n p_i \cdot L_i
	\end{equation}
	where $L(.)$ is the loss function for each participant's learning task, $W$ is the model parameters, and $p_i$ is the weight to represent each node's importance. In general, $p_i$ is decided by considering the node's data set size $|D_i|$ so that each instance, regardless of the location or data owner, has equal importance contributing to the overall loss. Sometimes, we use $L_i$ as a brief denotation of the $L(D_i,W)$. 
	
\section{Statistical heterogeneity in federated learning} \label{seq:stat-hete}
	
\begin{figure*}
	\centering
	\includegraphics[width=0.85\textwidth]{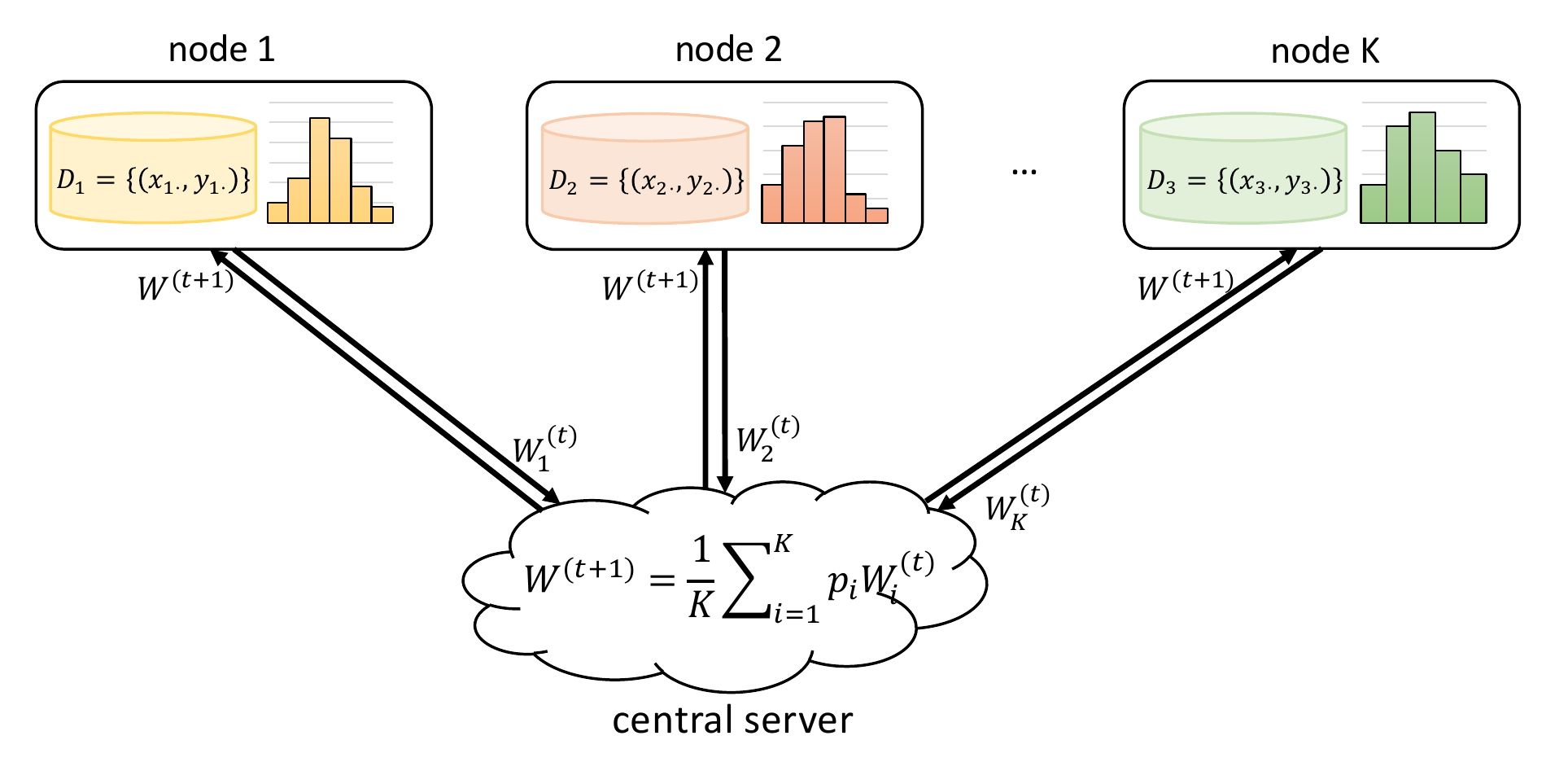}
    \caption{Statistical heterogeneity problem in federated learning}
    \label{fig:data-hete}       
    \end{figure*}

	One challenge of federated learning is statistical heterogeneity in which users have different data distribution. Statistical heterogeneity is an inherent characteristic of a user's behaviour. It is also identified as a non-IID problem. Conventional machine learning is built upon the IID assumption of a uniform dataset. The stochastic gradient descent (SGD) optimization used in vanilla federated learning is not specifically designed and optimized for tackling non-IID data. As described in Fig. \ref{fig:data-hete}, each participant's data are generated from different distributions. Each local model should then be initialized by the global model that represents a particular distribution.
	
	From the Bayes theorem perspectives, the classifications are highly linked to several distributions: $p(x)$, $p(y)$, $p(x|y)$ and $p(y|x)$. The variance of any distribution across participants will cause inconsistency in learning that will eventually damage the performance of a federated learning task. To solve this challenge, the simplest solution is to enhance existing distributed machine learning to be more robust for tolerating heterogeneity across data nodes \cite{li2018federated}, or to re-weight the contribution of each node according to the similarity of global and local models \cite{jiang2020decentralized}. However, that solution still uses one global model to service all variety of participants. A better solution is to increase the number of global models from one to multiple (Section \ref{seq:cluster-fl}). We can cluster or group the participants so that same-group participants with the same or similar distributions will contribute one of the global models, and each group will have a unique global model. Personalized federated learning (Section \ref{seq:per-fl}) is a recent method that tries to increase the number of global models to be equivalent to the number of participants, and then each participant will have a unique model including both a commonly shared part and personalized information.
	
	\subsection{Clustered federated learning} \label{seq:cluster-fl}
	A machine learning model could be treated as a function to approximate the distribution. In general, two models with similar parameters or functions are more likely to produce a similar outcome regarding the same input. Therefore, measuring models' similarity is an indirect way to measure distribution. Below are the related methods for clustered federated learning.
	
	\begin{table*}[htbp!]\small
        \begin{center}
	    \caption{Comparison of clustering-based federated learning}
	    \begin{tabular}{||m{2.5cm}|m{2cm}|m{1.8cm}|m{2cm}|m{2.5cm}||}
	        \hline
	        Methods & Motivation & Clustering & Factors & Measurement   \\
	        \hline\hline
	         Multi-center FL\cite{xie2020multi} & Better initialisation & K-means & Model parameters & L2-distance \\
	         \hline
	         Hierarchical clustering-based FL \cite{sattler2019clustered,briggs2020federated} & Similar distribution & Hierarchical clustering  & Gradients & Cosine similarity \& L1/L2 distance  \\
	         \hline
	         Hypothesis clustering-based FL \cite{mansour2020three,ghosh2020efficient} & Better hypothesis & K-means & Test accuracy & The loss of hypothesis  \\
	         \hline
	    \end{tabular}
	    \label{tab:my_label}
        \end{center}
	\end{table*}
	
	Xie \textit{et al.} \cite{xie2020multi} addresses the non-IID challenge of federated learning and proposes a multi-center aggregation approach. The non-IID problem in federated learning is defined as a joint optimization problem of multiple centers/clusters across participants. It can simultaneously learn multiple global models from participants with non-IID data, and the clustering and model learning are jointly optimized in a stochastic expectation maximization framework. 
	In particular, the loss function of the federated learning framework is defined as below.
	\begin{equation}
	    \sum_{i=1}^n p_i \cdot \min_{k} ||W_i - W^{(k)}||^2
 	\end{equation}
 	where the similarity of two models is measured by the L2 distance between the $i$-th participant's model $W_i$ and the global model $W^{(k)}$ of the cluster $k$.
    
    In some cases using Convolutional Neural Networks (CNN) as basic model architecture, two neurons from different models may have similar functions but with different neuron indices. Therefore, neuron matching of two CNN models cannot be simply applied to index-based matching, and it needs to be carefully considered for functionality matching. A proper neuron matching mechanism in the context of federated learning can improve the performance \cite{wang2020federated}. It could be further applied to clustering-based federated learning that considers matching neurons in both averaging and clustering steps.
    
    \cite{sattler2019clustered} proposes to distinguish participants based on their hidden data generating distribution by inspecting the cosine similarity $\alpha_{i,j}$ between their gradient updates $r_i$ and $r_j$. Based on the measurement in Eq. \ref{eq-distance-hyper}, a hierarchical clustering method is proposed to iteratively split participants into multiple groups, in which pairs of participants with larger similarity are more likely to be allocated to the same group. Below is the equation to calculate the similarity between a pair of participants $i$ and $j$. Moreover, \cite{briggs2020federated} discussed using different measurements.
    \begin{equation} \label{eq-distance-hyper}
        \alpha_{i,j} := \alpha(\nabla r_i(W_i), \nabla r_j(W_j)) 
        := \frac{<\nabla r_i(W_i), \nabla r_j(W_j)>}{||\nabla r_i(W_i)|| \;  ||\nabla r_j(W_j)||}
    \end{equation}
    where the $W_i$ and $W_j$ are the model parameters of participants $i$ ad $j$ respectively. 
    
	Mansour \textit{et al.} in \cite{mansour2020three} use a performance indicator to decide the cluster assignment for each node. In particular, given $K$ clusters with model $F_k$ controlled by $W$, the participant $i$ will be assigned to cluster $k$ whose model will generate the minimal loss $L$ using the test data set $D_i$ from participant $i$. The overall loss can be rewritten as follows.
	\begin{equation}
	    \sum_{i=1}^n p_i \cdot \min_{k} \{ L(D_i,W_k) \}
	\end{equation}
	in where $W_{(k)}$ is the parameters of the $k$-th global model/hypothesis.
	The paper gives a comprehensive theoretical analysis of the given method. Then, \cite{ghosh2020efficient} conducts convergence rate analysis in the same method. \cite{mohri2019agnostic} proposes a similar solution from a mixture of distribution perspectives.

\subsection{Personalized modelling} \label{seq:per-fl}
	When a service provider wants to provide a service that is the best for each individual customer, the model trained in the central server needs to be personalized or customized. The simplest solution is to treat the global model as a pre-trained model, and then use local data to fine-tune the global model, which will derive a personalized model. However, in most cases, each participant just has a limited number of instances, and the fine-tuning operation will cause over-fitting or increase the generalization error. Another solution is to treat each customer as a target task and the pre-trained global model as a source task, then to apply transfer learning \cite{pan2009survey} or domain adaptation \cite{blitzer2008learning} methods to fine-tune each personalized model. These methods will further leverage the global information to improve the fine-tuning process for each participant. \cite{mansour2020three} discusses two approaches, namely Data Interpolation and Model Interpolation, to learn a personalized federated learning model by weighting two components between local and global servers in terms of data distributions or models respectively. 
	
	\paragraph{Personalization layers} In general, a model can be decomposed into two parts: a representation learning part and a decisive part. For example, CNN is composed of convolution layers for representation extraction and fully-connected layers for classification decision. In a federated learning setting, heterogeneity could impact one of the two parts. \cite{arivazhagan2019federated} proposes to share representation layers across participants, and then keep decision layers as a personalized part. \cite{liang2020think} thinks representation layers should be the personalized part, and then the decision layers could be shared across participants. 
	
	\paragraph{Mixture models} If we cannot clearly identify which parts in the model will be impacted by the heterogeneity, we can roughly mix the global model and local models to incorporate both common knowledge and personalized knowledge. \cite{deng2020adaptive} proposes a mixture model of global and local models in a federated setting. The local model can preserve the personalized information and the global model provides common information. In particular, the loss is designed as below. 
	\begin{equation}
	    \sum_i^n p_i \cdot \{ L(D_i, W) + \lambda L(D_i, W_i) \}
	\end{equation}
	where $\lambda$ is the mixing weight, $L(D_i, W)$ and $L(D_i, W_i)$ are the loss of participant $i$ using global and local model's parameters $W$ and $W_i$ respectively. 
	
	\paragraph{Personalized models with constraints} In FedAVG, the pre-trained global model will be deployed to each participant for direct use. However, each participant could use their own data to fine-tune the pre-trained global model to generate a personalized model. \cite{dinh2020personalized,hanzely2020federated} enables participants to pursue their personalized models with different directions, but use a regularization term to limit each personalized model to not far away from the “initial point”, the global model. A regularization term will be applied to each participant's personalized model to limit the distance of personalized changes. The model is to be optimized using the loss function below.
	\begin{equation}
	    \sum_{i=1}^n p_i \cdot \{ L(D_i, W_i) + \frac{\lambda}{2} ||W_i - W||^2 \}
	\end{equation}
	where $L$ is the loss function decided by dataset $D_i$ and the $i$-th participant's model parameter, and $W$ is the global model's parameter. The regularization term could be attached to the global model or local model respectively. Moreover, it could be added to the federated learning process \cite{li2018federated}.
	
	\cite{zhao2018federated} allows each participant to take one more gradient descent step from the global model. This one step optimization is toward a personalized model. The loss is changed as below.
	\begin{equation}
	    \sum_i^n p_i \cdot L_i(W-\nabla L_i(W))
	\end{equation}
	where $L_i$ is the loss function of the $i$-th participant that is controlled by the weights $W$ and data set $D_i$.

\section{Model heterogeneity} \label{seq:model-hete}

    \subsection{Model architecture heterogeneity}
\begin{figure*}
	\centering
	\includegraphics[width=0.85\textwidth]{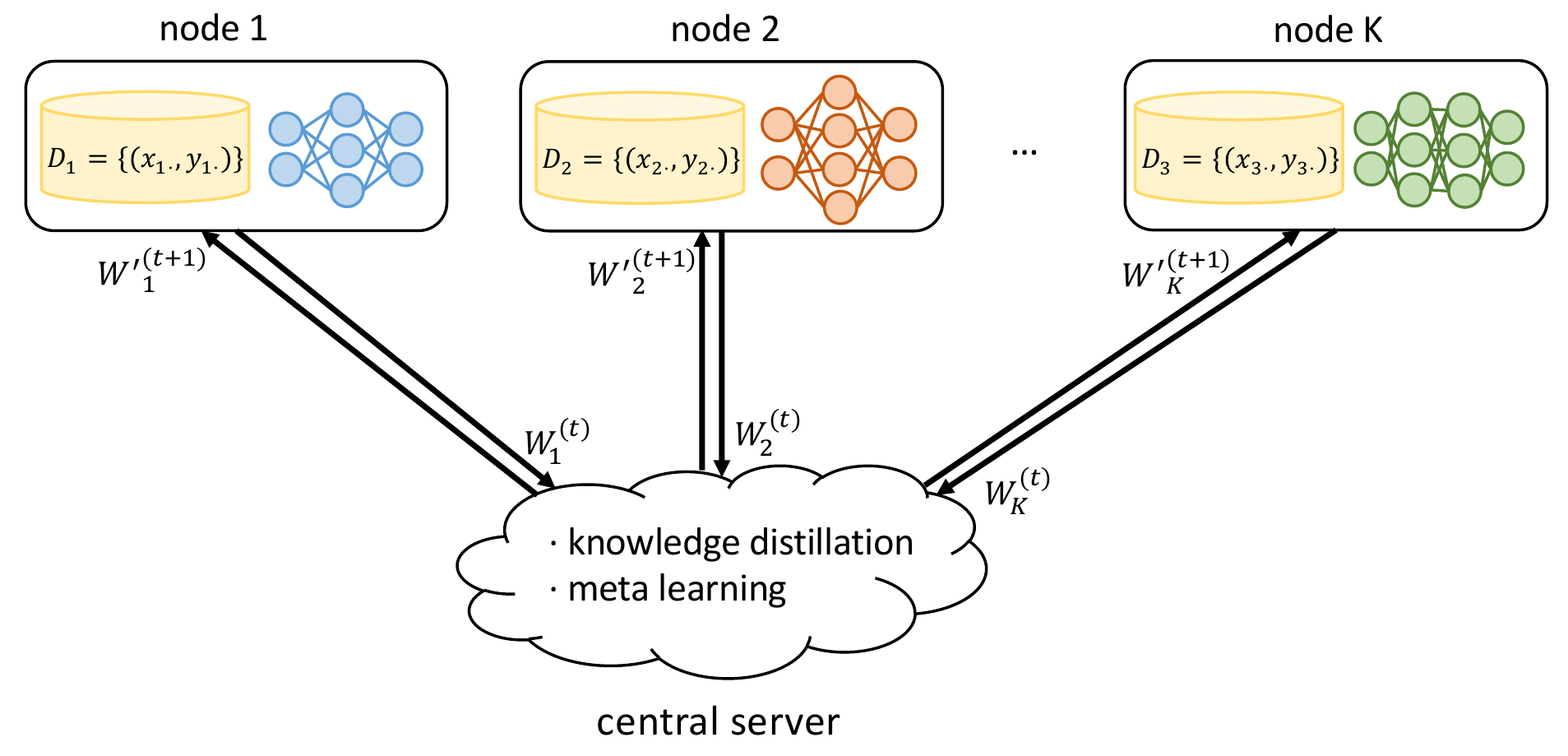}
    \caption{Mode heterogeneity problem in federated learning}
    \label{fig:model-hete}       
    \end{figure*}

	Model heterogeneity will ruin the model aggregation operator that is the core part of federated learning. To enable the aggregation, we need to find a way to transform heterogeneous models into homogeneous models. As shown in Fig. \ref{fig:model-hete}, the participants in a federated learning system could have heterogeneous models with different architectures. Therefore, the global model $W'$ is also different from each local model $W$. It will become a challenge to aggregate the heterogeneous model in the federated setting.
	
	Knowledge distillation is such a technology to ``compress" or ``distill" a large model into a small model. It was proposed by Hinton \textit{et al.} \cite{hinton2015distilling} in 2015. It can extract information from a ``teacher" model $W'$ into a simpler ``student" model $W$. Given the same inputs, the objective function is to control the student model to produce similar outputs (probability vector) with the teacher model while considering ground truth. The loss function is defined below.
	
	\begin{equation}
	    L_s(\{(x,y')\}, W) + \lambda \cdot L_h(\{(x,\hat{y})\},W)
	\end{equation}
	where $\hat{y}$ and $y'$ represent the label from ground truth and teacher model's predicted probability of labels, $W$ is the parameters for the student model, $L_s$ is a soft loss function to measure the dissimilarity or distance between two distributions of predicted labels from teacher model and student model respectively, $L_h$ is a loss function between predicted labels and ground truth, and $\lambda$ is an importance weight of $L_s$ and $L_h$.
	
	In \cite{li2019fedmd}, each participant owns a private dataset and a public dataset is shared across all participants and servers. A local model is trained in a transfer learning framework by leveraging both public and private datasets. Then, a decentralized knowledge distillation will be applied to exchange information between participants and the coordinator. In particular, each participant calculates the prediction probability for all instances in the public dataset, and then sends it to the coordinator for prediction averaging. The averaged prediction probability could be viewed as the output of a teacher model (teacher) that is an ensemble learning of many local models, and the teacher model's outputs will be sent back to each participant to train a student model in a knowledge distillation process. In this work, the heterogeneous local models exchange prediction probability acquired on the public dataset, and knowledge distillation allows exchanging information in a model agnostic way.
	
	The authors in \cite{guha2019one} leverage distillation in a semi-supervised setting to reduce the size of the global model. The distillation mechanism also potentially adds privacy guarantees by replacing some sensitive model parameters during the distillation process. In \cite{jeong2018communication}, a new federated distillation (FD) is proposed to improve communication efficiency for an on-device machine learning framework. The proposed FD exchanges the model output rather than the model parameters, which allows the on-device ML to adopt large local models. Prior to operating FD, the non-IID datasets are rectified via federated augmentation, where a generative adversarial network is used for data augmentation under the trade-off between privacy leakage and communication overhead.
	
	This technique can be used to bridge the gap between a global model and a local model for a specific client. In \cite{yu2020salvaging}, the pre-trained global model is treated as the teacher, while the adapted model is treated as a student. The participant will train a unique local model with specific architecture that is different from the others. However, this solution assumes the federated model is pre-trained, and the knowledge distillation is just for the deployment stage or personalization step rather than federated learning.

\section{Limited number of uses of participants} \label{seq:access-limit}
    In an open banking data marketplaces, the use of participants' data may be charged by times. For example, in a federated learning process, if the coordinator asks the participant to train the local model three times, they should be charged by three times as well. Moreover, the participant's data is a dynamically changing profile including its banking-related activities. To capture the dynamic changes of the customers, the federated learning may take an incremental or lifelong learning strategy to regularly use a participant's data to refine the global model. This will bring a new challenge for continuously sharing data in a long-term model training framework.
	
	The pay-per-access mechanism can also be modeled as a few-shot federated learning problem in which the server can take very few communication rounds with each participant. Regardless of whether the federated learning is horizontal or vertical, the participants can help the coordinator to train a local model according to the global model's parameters at a particular moment. Thus, the coordinator will pursue highly efficient communication with participants using their data.
	
	Some techniques can be used to solve the few-shot federated learning challenge in open banking. In \cite{guha2019one}, Guha \textit{et al.} present one-shot federated learning that allows the central server to learn a global model over the network in a single round of communication. The proposed approach utilizes ensemble learning and knowledge aggregation to effectively and efficiently leverage information in large networks with thousands of devices. With this technique, the financial service provider can access the data from multiple data holders and finish model training in only one round of communication. This greatly improves the efficiency of communication and reduces the risk of sensitive data leakage.

\section{Only positive labels in each participant} \label{seq:one-class}
\begin{figure*}
	\centering
	\includegraphics[width=0.85\textwidth]{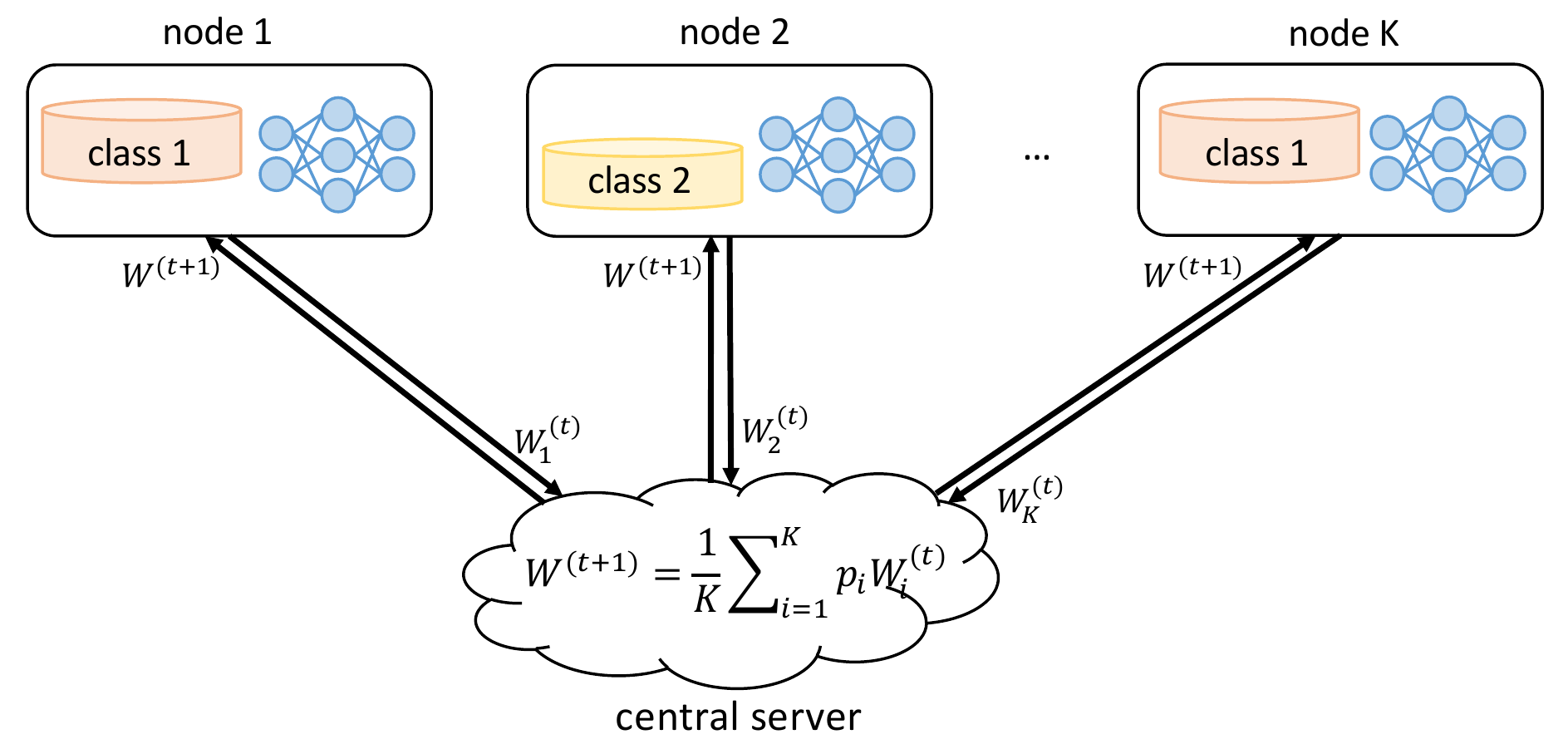}
    \caption{One-class problem in federated learning}
    \label{fig:one-class}       
    \end{figure*}

	The one-class challenge is also a problem. For example, in a fraud detection or break contract, one user’s data can only be labeled as fraud or not-fraud. Most users are labeled as the not-fraud class, which means that there are only a small number of users with the fraud class. Although each user can design and train personalized model based on its own financial data, the model may not be accurate enough as a result of the one-class problem. As shown in Fig. \ref{fig:one-class}, the overall learning task is a bi-class problem although each participant has only a one-class learning task. The participant cannot properly train a bi-class classifier to be aggregated in the coordinator's model or global server.
	
	In open banking scenarios, solutions to the one-class problem can fall into two categories. One is to embed the specific one-class classification algorithm into the imbalanced federated financial system from a task-level perspective. The other is to adjust the weights and incentive mechanism among users from a system-level perspective.
	One-class classification is also known as anomaly detection in fraud detection applications, where outliers or anomalies are rare examples that do not fit in with the rest of the data. 
	
	In \cite{yu2018federated}, the authors propose a federated learning-based proactive content caching (FPCC) scheme which is based on a hierarchical architecture where the server aggregates user-side updates to construct a global model. Based on a stacked auto-encoder, FPCC allows each user to perform training on its local data using hybrid filtering. In open banking, the users' similarity in a financial perspective can be calculated using the features generated from a stacked auto-encoder. In this way, it is easier to make fraud detection with imbalanced data distribution. 
	
	The authors in \cite{preuveneers2018chained} addressed that federated learning setup allows an adversary to poison at least one of the local models and influence the outcome of one-class classification. Moreover, malicious behaviour can increase time-to-detection of the attacks. To solve these above issues, they designed a permitted blockchain-based federated learning method where the updates for the auto-encoder model are chained together on the distributed ledger. The trained auto-encoder can recognize the test samples of the baseline class while marking the other test samples that do not fit into the trained model as a potential negative class.

\section{Summary}
    This chapter discusses the challenges of applying federated learning in the context of open banking. In particular, we focus on discussing the statistical heterogeneity (Section \ref{seq:stat-hete}), model heterogeneity (Section \ref{seq:model-hete}), access limits (Section \ref{seq:access-limit}), and one-class problems (Section \ref{seq:one-class}) that are rarely discussed in other places. This chapter explores various solutions to solve the aforementioned practical challenges in open banking, and then foresee the advancement of federated learning in the context of real-world scenarios.
	
\bibliographystyle{splncs04}
\bibliography{FL_book}
	
\end{document}